\documentstyle[preprint,prc,aps,epsfig]{revtex}
\tighten
\begin{document}
\bibliographystyle{revtex}
\draft
\title{Discerning the neutron density distribution of $^{208}$Pb
from nucleon elastic scattering}
\author{S. Karataglidis$^1$, K. Amos$^{1,2}$, B. A. Brown$^3$, and
P. K. Deb$^2$}
\address{$^1$Theoretical Division, Los Alamos National Laboratory, Los
Alamos, New Mexico, 87545\\ $^2$School of Physics, University of
Melbourne, Victoria 3010, Australia \\ $^3$National
Superconducting Cyclotron Laboratory and Department of Physics and
Astronomy, Michigan State University, East Lansing, Michigan,
48824-1321}
\preprint{LA-UR-01-6133}
\date{\today}
\maketitle
\begin{abstract}
We seek a measure of the neutron density of $^{208}$Pb from analyses
of intermediate energy nucleon elastic scattering. The pertinent model
for such analyses is based on coordinate space nonlocal optical
potentials obtained from model nuclear ground state densities. Those
potentials give predictions of integral observables and of angular
distributions which show sensitivity to the neutron density. When
compared with experiment, and correlated with analyses of electron
scattering data, the results suggest that $^{208}$Pb has a
neutron skin thickness $\sim 0.17$~fm.
\end{abstract}

\section{Introduction}
Interest in the matter distributions of $^{208}$Pb, and its neutron
density profile particularly, is quite topical \cite{Br00}. There is a
proposal to measure its neutron root-mean-square (rms) radius at the
Jefferson Laboratory \cite{Je00} from an analysis of parity-violating
electron scattering data. In contrast to proton rms radii that are
known to within an accuracy $\sim 0.02$~fm \cite{Fr95}, neutron rms
radii are less certain.

Recently, the neutron rms radius in $^{208}$Pb was assessed in terms
of modern Skyrme-Hartree-Fock (SHF) models \cite{Br00}. With the
Friedman-Pandharipande (FP) neutron equation of state \cite{Fr81} as a
constraint, the neutron rms radius in $^{208}$Pb was expected to be
$0.16 \pm 0.02$~fm larger than the proton value. Previous estimates of
this neutron skin, $S = \sqrt{ \left\langle r^2_n \right\rangle } -
\sqrt{ \left\langle r^2_p \right\rangle }$, ranged from 0.1~fm to
0.3~fm \cite{Je00,Po97}; the lower values favored in general by SHF
models, while relativistic mean field models predict values closer to
0.3~fm \cite{Ty01}. Knowledge of the skin thickness in $^{208}$Pb then
is a good constraint upon such models of structure \cite{Ty01}.

The planned parity-violating electron scattering experiment
\cite{Je00} will only provide information about the neutron rms radius
itself. We seek further information and address the question of
whether analyses of nucleon scattering data establish a measure of the
neutron density distribution in $^{208}$Pb. Hadron scattering data
have been analyzed previously to deduce the neutron skin thickness in
$^{208}$Pb. The ratio of $\pi^+$ and $\pi^-$ reaction cross sections
gave $S = 0.0 \pm 0.1$~fm \cite{Al73}. Analyses of elastic proton
scattering data at 0.8~GeV gave $0.14 \pm 0.04$~fm
\cite{Ho80}. However, a review of the analysis of proton scattering
from $^{40}$Ca \cite{Ra92} gave values in the range of $-$0.4~fm to
$-$0.2~fm for $S$ in that nucleus which are systematically smaller
than all theoretical models which give $-$0.05~fm \cite{St82}, and
suggests that there is a systematic problem in the analysis using
phenomenological models of high energy proton scattering data which
would affect the extracted $S$ values at the level of $\sim 0.2$~fm. A
more recent analysis of 650~MeV proton scattering data \cite{St94}
gave $S = 0.20 \pm 0.04$~fm for $^{208}$Pb while that reaction model
gave a result for $^{40}$Ca which is consistent with theoretical
predictions. The excitation of the isovector giant dipole resonance in
$^{208}$Pb by inelastic alpha scattering \cite{Kr94} was used to
deduce $S = 0.19 \pm 0.09$~fm. These previous analyses, based on
phenomenological models, produce a range of results for $^{208}$Pb and
suggest that there may be small but systematic errors in those strong
interaction models for hadron scattering limiting the accuracy in the
extraction of $S$ to $\sim 0.2$~fm.

Our approach is based on coordinate space nonlocal optical potentials
generated by a full folding of realistic effective nucleon-nucleon
($NN$) interactions with ground state density matrices (termed
densities hereafter) of $^{208}$Pb. This allows us to distinguish
between various model structures, including those of the SHF type
proposed by Brown \cite{Br00}, even if two may have the same rms
radii. As a calibration of the use of SHF models we consider the
elastic scattering from $^{40}$Ca as well.  Further, as the effective
$NN$ interaction is dominated by the isoscalar $^3S_1$ channel
\cite{Review}, proton scattering predominantly will probe the neutron
density and {\em vice-versa} and so we consider both proton and
neutron elastic scattering at a given energy seeking as complete a map
as possible of the nuclear matter distributions.

\section{Nuclear models and the microscopic optical potential}

The model with which predictions of nucleon-nucleus ($NA$) scattering
observables are made has been given in detail in a recent review
\cite{Review}. Use of the complex, nonlocal, $NA$ optical potentials
defined by that model prescription, without localization of the
exchange amplitudes, gave predictions of differential cross sections
and spin observables that are in good agreement with data from many
nuclei ($^3$He to $^{238}$U) and for a wide range of energies (40 to
300~MeV).  Crucial to that success was the use of effective $NN$
interactions built upon $NN$ $g$ matrices; solutions of
Brueckner-Bethe-Goldstone equations for realistic starting (free) $NN$
interactions.  The $NA$ optical potentials result from folding those
effective interactions with the densities of the target nucleus. That
folding includes the antisymmetrization of the projectile-nucleus wave
functions and therefore exchange (knock-out) amplitudes are treated
explicitly.  Consequently the $NA$ potentials inherently are
non-local.  The optical potentials that result from that process we
term for brevity as $g$-folding potentials.

Recently, this approach was applied successfully to make predictions
of the integral observables of nucleon elastic scattering
\cite{De01}. Thus, the use of the $g$-folding optical potentials give
good predictions to both angular-dependent and integral observables; a
result not guaranteed with the more common phenomenological
approaches.  Of import, however, is that the level of agreement with
data in the $g$-folding model depends on the quality of the underlying
model of structure.  We seek to use that dependence as a sensitive
evaluation of the densities considered.

\begin{figure}
\centering\epsfig{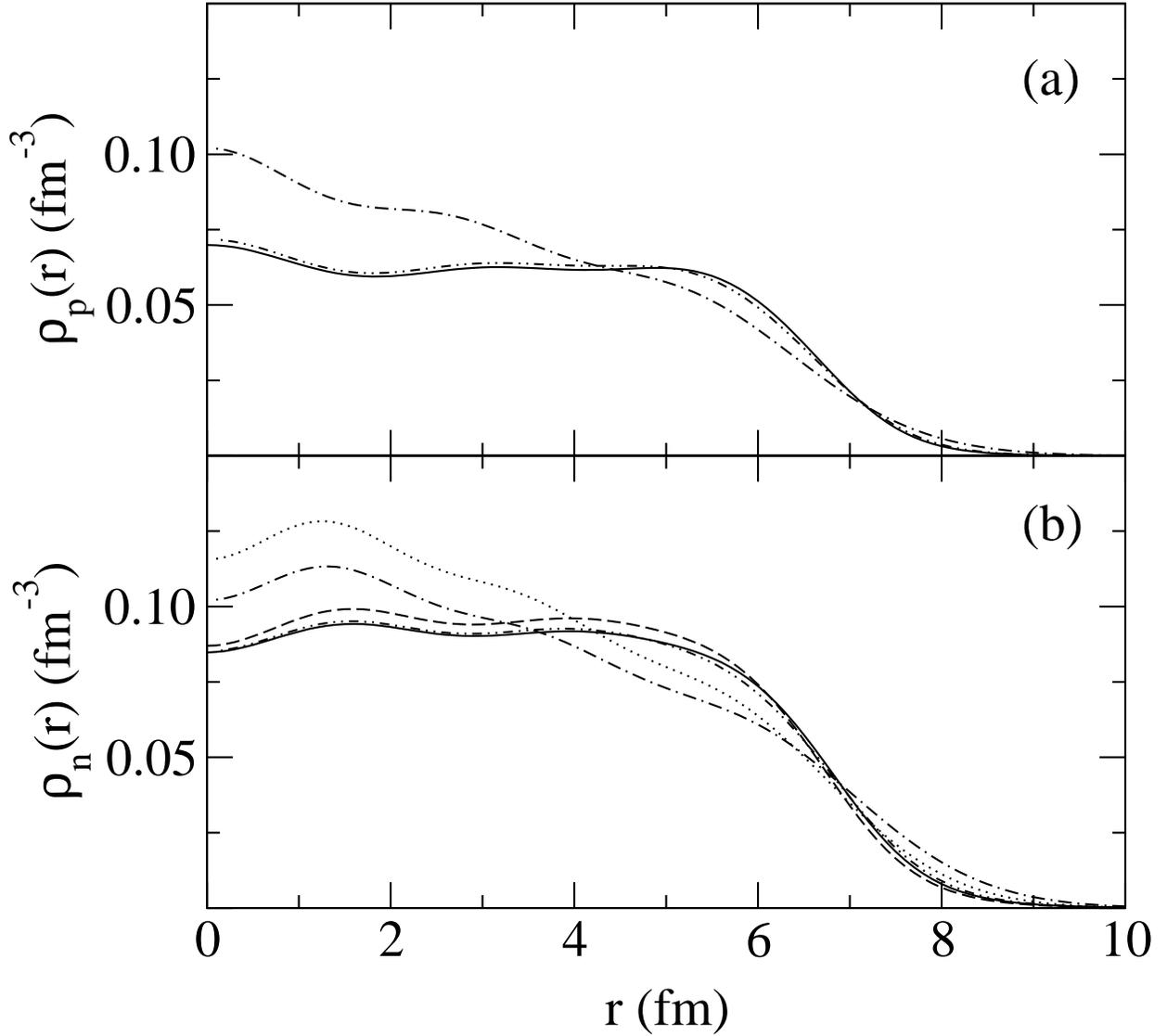}
\caption[]{Nucleon densities in $^{208}$Pb. The solid and dot-dashed
curves in the proton densities $\rho_p$ portray both SHF and both HO
models respectively. The neutron densities $\rho_n$ given by the
solid, dashed, dotted, and dot-dashed lines portray respectively the
SHF1, SHF2, HO1, and HO2 models. The double-dot-dashed line in each
case denotes the density obtain from the SKM* model.}
\label{densities}
\end{figure}

Our theoretical density distributions are based upon the Skyrme
Hartree-Fock model for $^{208}$Pb with a spherical closed-shell
configuration. It was shown in \cite{Br00} that there is a combination
of parameters in the Skyrme Hamiltonian (dominated by the $x_3$
parameter) which has a strong influence on the neutron skin thickness
but which are not well determined by the data on binding energies and
charge radii. This combination can be correlated with the pressure in
the neutron equation of state at normal nuclear density
\cite{Br00}. It can also be related to the surface symmetry energy
\cite{St82}. The Skyrme interaction SKX \cite{Br98} was obtained with
a constraint on the neutron equation of state provided by the FP model
\cite{Fr81} which constrains the neutron skin in $^{208}$Pb to be $S =
0.16 \pm 0.02$~fm. The results obtained with the SKX model will be
denoted herein by SHF1. If one does not allow for any constraint from
a model for the neutron equation of state then a much larger range of
$S$ is allowed. In particular, we use a model which is constrained by
the same nuclear properties used for SKX but with a value of the
parameters which gives $S = 0.25$~fm for $^{208}$Pb. This we denote as
SHF2. The densities for $^{208}$Pb obtained from these models, as well
as the others considered herein, are shown in Fig.~\ref{densities}.

\begin{figure}
\centering\epsfig{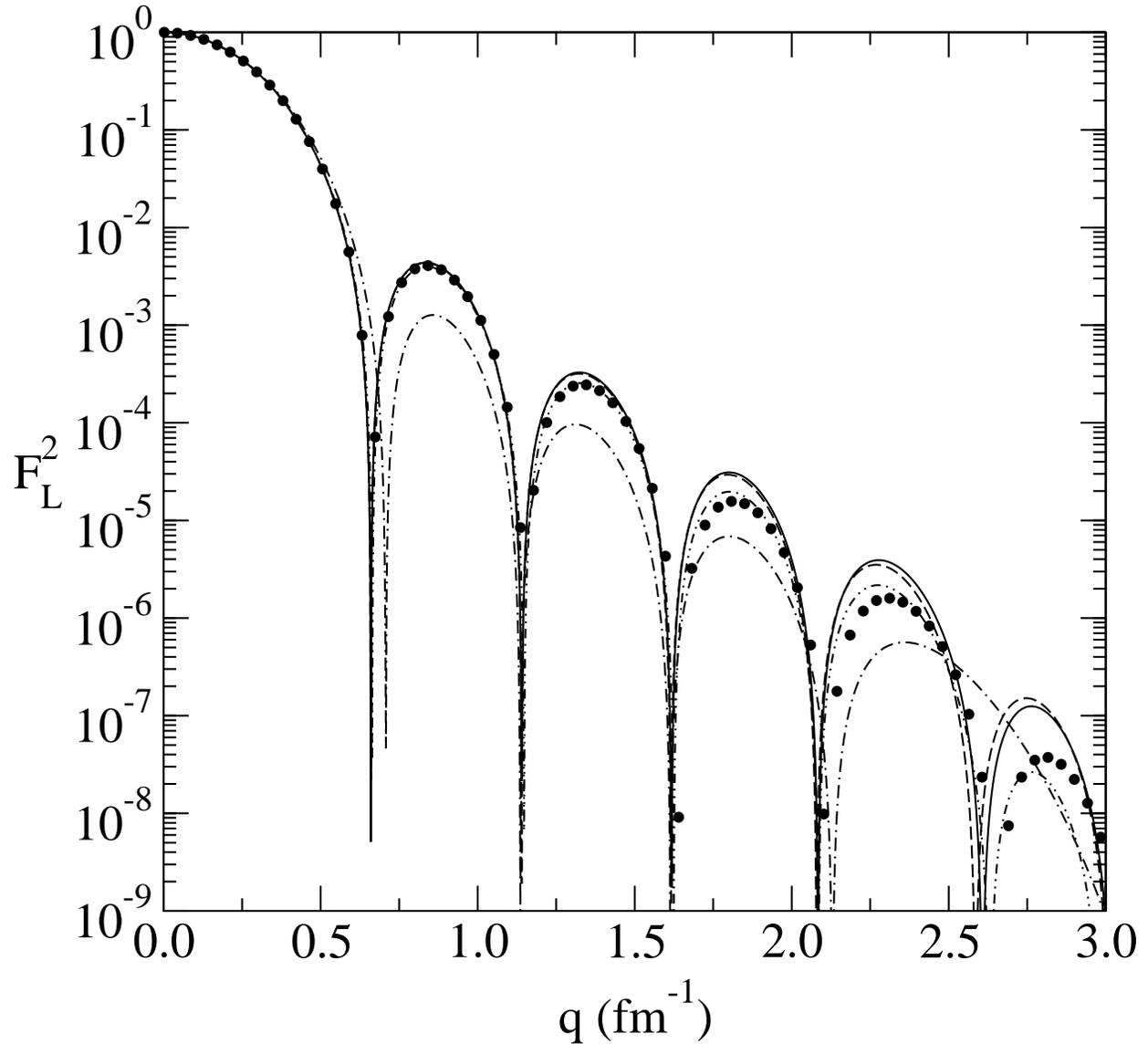}
\caption[]{Longitudinal elastic electron scattering form factor for
$^{208}$Pb. The data \cite{Vr87} are compared to the results of the
calculations made using the SHF1, SHF2, and SKM* models portrayed by
the solid, dashed, and double-dot-dashed lines, respectively. The
oscillator result is portrayed by the dot-dashed line.}
\label{electron}
\end{figure}

An initial test of these interactions is provided by the elastic
electron scattering data which yields information on the charge
density which, in turn, gives information on the proton density. SKX
appears to give an excellent reproduction of the charge-density
distribution (Fig.~11 of \cite{Br98}). However, there is some model
dependence in the extraction of the charge density from electron
scattering data. It is better to compare to a representation which is
more closely associated with the actual data -- this is the plane-wave
transform of the charge density shown in Fig.~\ref{electron}. The
experimental form factor is obtained from the charge density
distribution given in \cite{Fr95}. Experiment is compared with the
SHF1 and SHF2 model results showing a disagreement with data which
systematically increases with momentum transfer. The main feature of
the distribution which affects the high-$q$ behavior is the surface
thickness. As discussed in \cite{Br98}, the SKX interaction appears to
have a surface thickness which is a little sharper than that
determined from experiment. We have thus looked at other Skyrme
interactions in terms of the data in Fig.~\ref{electron} and find the
older SKM* interaction \cite{Ba82} is much better than others in this
regard. SKM* appears to achieve this improvement by a decrease in the
power of the density, $\rho^{\alpha}$, associated with the
density-dependent part of the interaction from its value $\alpha =
1/2$ for SKX to $\alpha = 1/6$ for SKM*. Coincidentally, SKM* predicts
a neutron skin of 0.17~fm which is essentially the same as that
predicted by SKX (SHF1 model). This is obtained mainly because the
$x_3$ parameter was set to zero by default, since it is not well
determined by nuclear data.

The value of $\alpha$ is also associated with the incompressibility
coefficient $K$ for infinite nuclear matter which ranges from $K =
270$~MeV for SKX to $K = 217$~MeV for SKM*. It was found in \cite{Br98}
that $\alpha = 1/2$ gave the best overall fit with the data set
considered and that when $\alpha$ is decreased (as in the SKXm
interaction, which has $\alpha = 1/3$) the overall $\chi^2$ increased
mainly because of an increase in the contribution from the
single-particle energies. It was also argued in \cite{Br98} that an
improvement in the surface properties may require an additional
parameter in the Skyrme Hamiltonian associated with the next order
$d$-wave term in the expansion in terms of the range of the $NN$
interaction. Thus the present models are not perfect but they are good
enough for a discussion of the effects of the neutron skin and the
surface thickness on the proton and neutron scattering data. We
compare results obtained with three Skyrme interactions SHF1 (SKX),
SHF2, and SKM* which will enable us to explore the effect of neutron
skin (a comparison of SHF1 and SHF2) with a fixed surface thickness,
and surface thickness (a comparison of SHF1 and SKM*) with a fixed
neutron skin.

We compare also with results obtained with the simple
harmonic-oscillator radial wave functions for $^{208}$Pb which were
used in \cite{Review}.  Two sets of oscillator parameters were
used. For HO1 we use $\hbar\omega = 6.70$~MeV for both protons and
neutrons which is chosen to give the rms charge radius of 5.50 fm (a
proton rms radius 5.45 fm). HO1 has a neutron rms radius of 5.84 fm.
For HO2 the oscillator parameter for neutrons was changed to 7.25~MeV
to decrease the rms neutron radius to 5.61 fm so that the neutron skin
$S=0.16$~fm is close to that obtained with SHF1 and SKM*.

The rms radii from all models of the ground state of $^{208}$Pb
considered are listed in Table~\ref{radii}. All five give essentially
the same radius for the protons but they vary considerably in the
radius for the neutrons. The difference between the neutron and proton
rms radii given in the last column emphasizes that spread. Note that
the HO2 model was chosen to give the same rms radii as those of the
SHF1 model. The neutron radius obtained from the SKM* model is similar
to those of the SHF1 and HO2 models.  However, as is evident in
Fig.~\ref{densities}, each model gives distinctive density
distributions.  The normalization used is such that their volume
integrals equate to the proton and neutron numbers of 82 and 126
respectively. In Fig.~\ref{densities}, the proton density $\rho_p(r)$
of both the SHF1 and SHF2 models are displayed by the solid curve. The
density obtained from the SKM* model is given by the double-dot-dashed
line and exhibits a larger diffuseness compared to the densities of
the other Skyrme models. Likewise, both the HO1 and HO2 models have
proton densities as given by the dot-dashed curve. These quite
distinct shapes nevertheless give the same proton rms radius. However,
they differ in the longitudinal electron scattering form factor
as shown in Fig.~\ref{electron}.

The five model neutron densities $\rho_n(r)$ are also shown in
Fig.~\ref{densities}. The SHF1 and SHF2 neutron densities are
displayed by the solid and dashed lines respectively. They are similar
with the SHF2 density having a slightly more diffuse surface region; a
property that results in the larger neutron rms radius. The density
from the SKM* model is given by the double-dot-dashed line and, as for
the proton density, exhibits a larger diffuseness compared to the
densities of the other Skyrme models. The neutron densities of the
HO1 and HO2 models are displayed by the
dotted and dot-dashed lines respectively. As with their proton
densities, the neutron densities of both of these models are enhanced
in the nuclear interior over the SHF values. But these HO
densities also have increased neutron probability at the
surface. Recall that the HO2 model was set to have the
same neutron rms radius as the SHF1 prescription.

\begin{figure}
\centering\epsfig{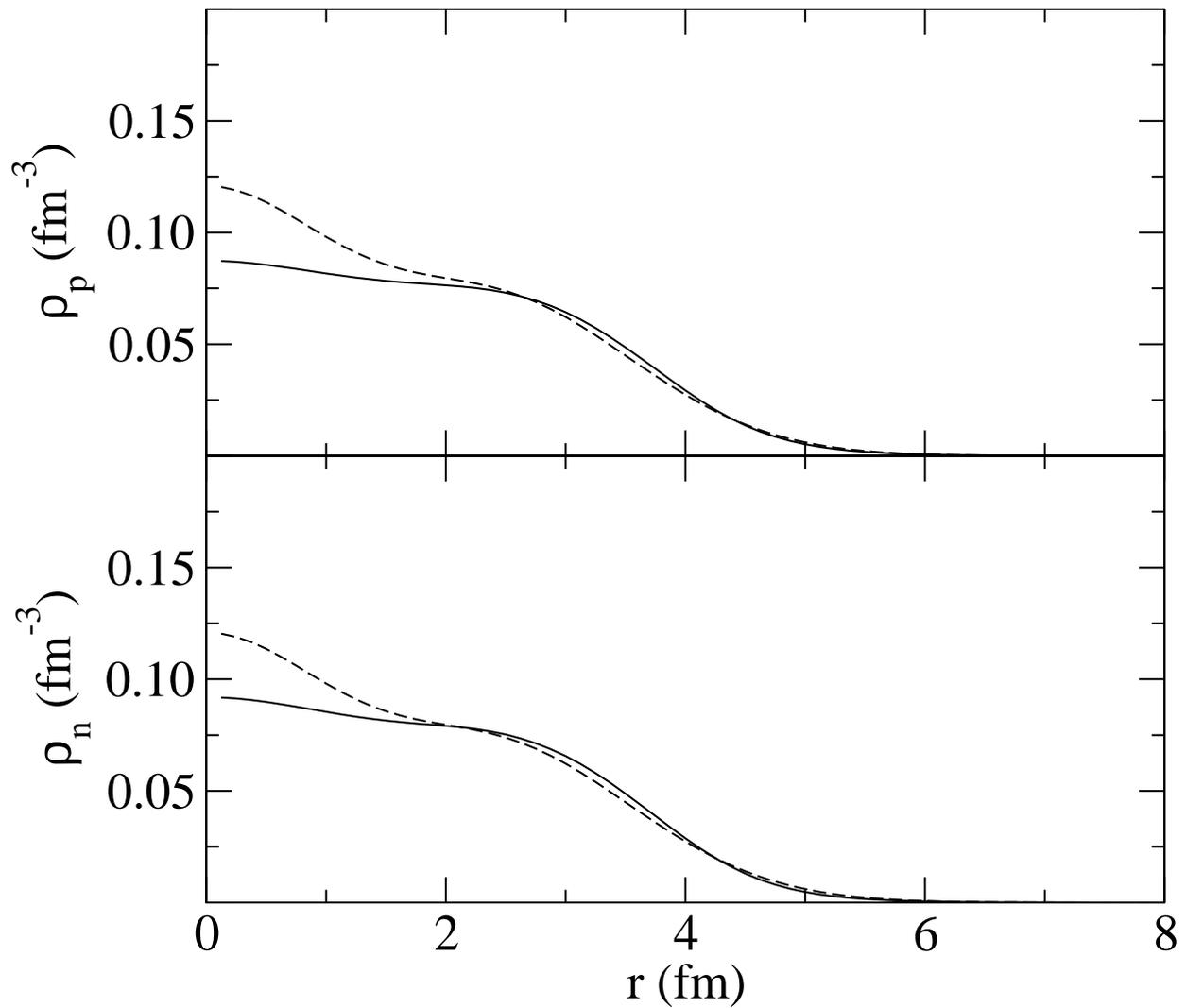}
\caption[]{Nucleon densities in $^{40}$Ca. The solid and dashed lines
portray the SHF and HO models respectively.}
\label{ca_dens}
\end{figure}

The SHF (SKX) densities for $^{40}$Ca are displayed in
Fig.~\ref{ca_dens}. Also displayed therein are the densities from the
oscillator model used by Karataglidis and Chadwick \cite{Ka01}, for
which $\hbar\omega = 10.25$~MeV. In the surface region, the two models
predict essentially the same densities.  As those densities differ
markedly only well within the nuclear volume, we expect influence on
scattering primarily at high momentum transfer scattering results at
energies for which absorption through the nucleus is not too large. We
anticipate differences in cross sections for scattering at energies
$\ge 200$~MeV and at momentum transfer values $\ge 1$~fm$^{-1}$.

\section{Results}
We have analyzed both proton and (where available) neutron elastic
scattering data from $^{40}$Ca and $^{208}$Pb at 40, 65, and
200~MeV. The choices of energies were predicated in part on the
availability of data and of the momentum transfer values at which those
data have been measured.  In addition our choice was influenced by
previous applications of the $g$-folding potentials with those
energies being quite successful~\cite{Review}.  Furthermore, the
effective interactions defined for each energy are quite different due
to the energy dependence of medium effects in the basic $g$ matrices
so that the set of $NA$ scattering results we obtain provide a
consistency check on the various models of structure used.

\begin{figure}
\centering\epsfig{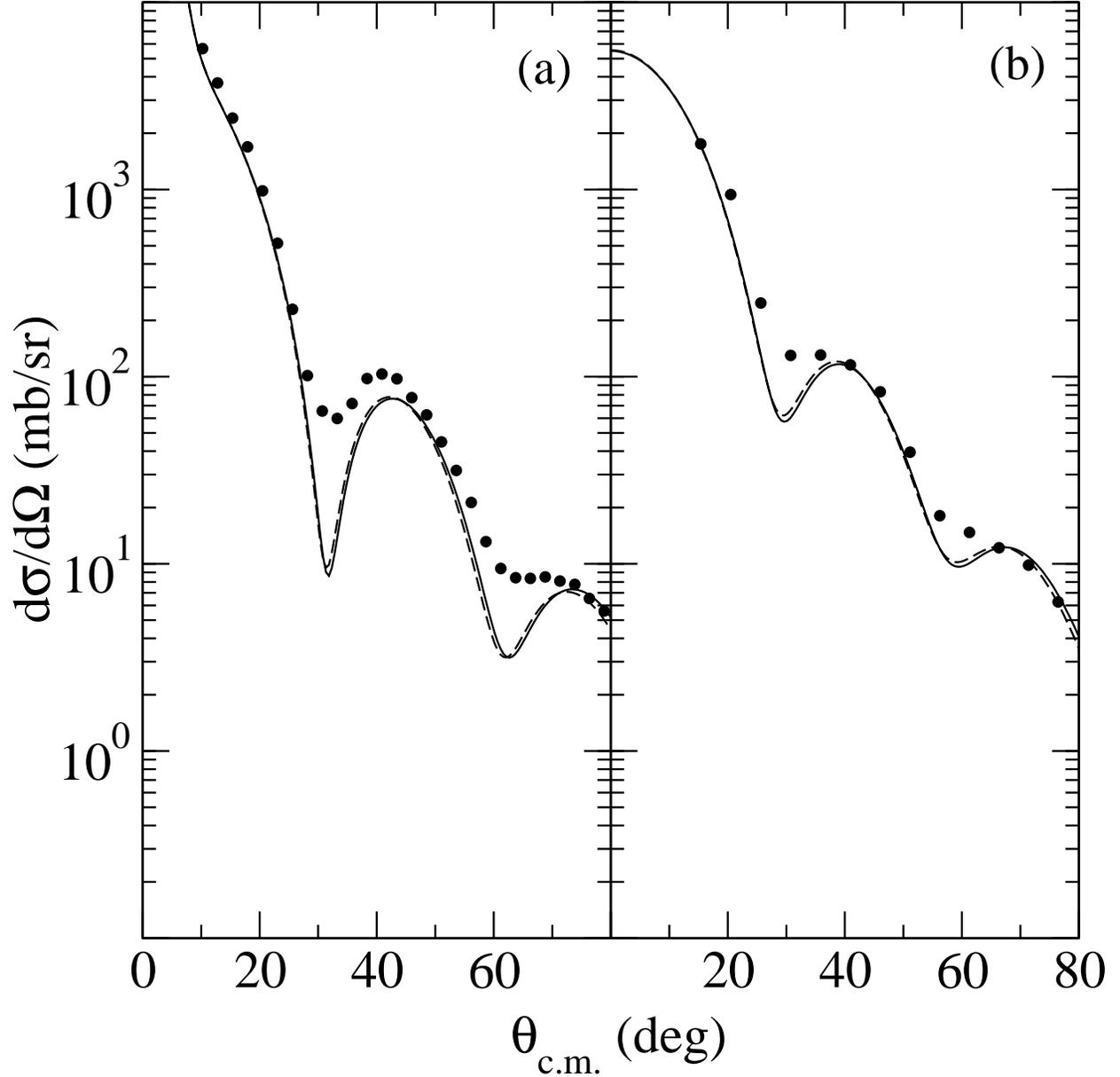}
\caption[]{Differential cross sections for 40~MeV nucleon elastic
scattering from $^{40}$Ca. The proton scattering data of Camis {\em et
al.} \cite{Ca86} are compared in (a) with the results of the
calculations made as defined in the text. In (b), the neutron
scattering data of de~Vito {\em et al.} \cite{Vi81} are compared
with the results of equivalent calculations.}
\label{ca_xsec_40}
\end{figure}

%
%	Ca diagrams first
%
The differential cross sections for 40~MeV proton and neutron elastic
scattering from $^{40}$Ca are presented in
Fig.~\ref{ca_xsec_40}. Therein, the proton scattering data of Camis
{\em et al.} \cite{Ca86} and the neutron scattering data of de~Vito
{\em et al.} \cite{Vi81} are compared to the results of the
calculations made using SHF (solid line) and HO (dashed line)
models. The SHF and HO model results equally well describe the data,
although they underpredict the proton scattering in the regions of the
minima. That disagreement may be due to problems in specifying the
effective interaction at low energies \cite{De00}. However, further
comment should await the consideration of the comparisons of 40 MeV
proton scattering from ${}^{208}$Pb.  Whatever any deficiency at this
energy there may be though does not affect results we find for, and
conclusions we may draw from, scattering at 65 and 200 MeV.

\begin{figure}
\centering\epsfig{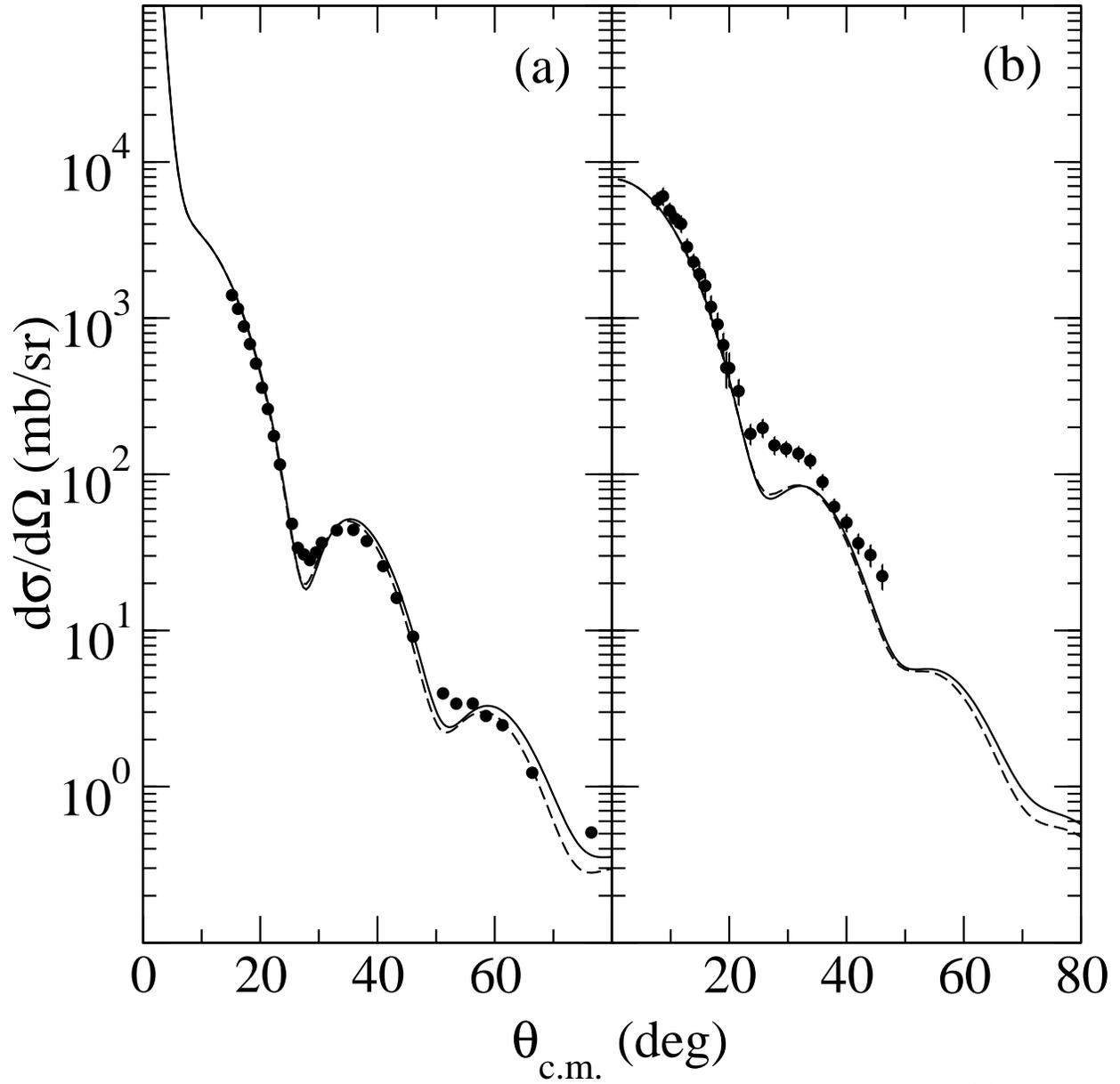}
\caption[]{As for Fig.~\ref{ca_xsec_40}, but for 65~MeV
scattering. The proton scattering data in (a) are those of Sakaguchi {\em et
al.} \cite{Sa82}, while the neutron scattering data in (b) are those
of Hjort {\em et al.} \cite{Hj94}.}
\label{ca_xsec_65}
\end{figure}

The 65~MeV elastic scattering scattering cross sections for $^{40}$Ca
are displayed in Fig.~\ref{ca_xsec_65}. Therein, the agreement between
the model results and the proton scattering data of Sakaguchi {\em et
al.} \cite{Sa82} is now very good; much better than that found in
Fig.~\ref{ca_xsec_40}.  Also, there is a slight preference for the SHF
result at larger angles. In the case of the neutron scattering
results, both model results agree but underpredict some of the data of
Hjort {\em et al.} \cite{Hj94}. However, those data are somewhat
problematic as concluded from a recent analysis \cite{Ka01} in which
several sets of data were compared with theory at that energy. A new
measurement of this cross section is required to resolve any such
problem.

\begin{figure}
\centering\epsfig{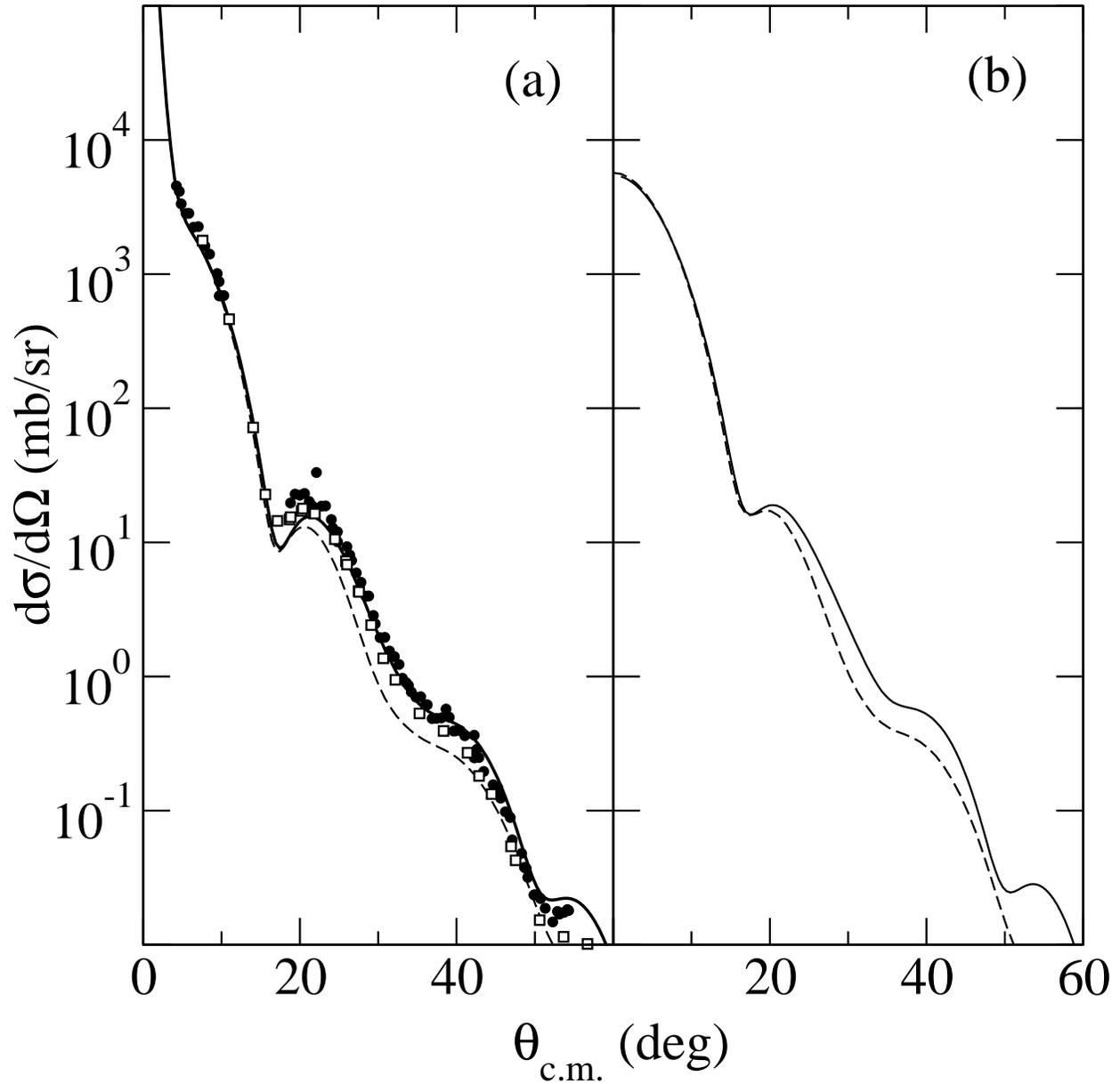}
\caption[]{As for Fig.~\ref{ca_xsec_40}, but for 200~MeV
scattering. The proton scattering data in (a) are those of Hutcheon
{\em et al.} \cite{Hu88} (circles) and Seifert {\em et al.} \cite{Se93}
(squares).}
\label{ca_xsec_200}
\end{figure}

It is with 200~MeV scattering from $^{40}$Ca, displayed in
Fig.~\ref{ca_xsec_200}, that we observe significant differences
between the predictions of scattering made using the SHF and HO
models.  For proton scattering that is displayed in
Fig.~\ref{ca_xsec_200}(a), the SHF model result agrees well with the
data of Hutcheon {\em et al.} \cite{Hu88} (circles) and Seifert {\em
et al.} \cite{Se93} (squares) and is a significant improvement on the
result found using the HO model.  This variation concurs with our
expectation that the differences between the inner radial densities of
the two models of structure would influence the results of scattering
calculations at higher energies. Taking the results at all three
energies, we believe that the SHF model is the better description of
$^{40}$Ca.

%
% 	Now for Pb
%
We now turn to $^{208}$Pb and consider first the integral observables
from nucleon scattering as a test of sensitivity to the matter
distributions of $^{208}$Pb.  Those integral observables at 40, 65,
and 200~MeV are given in Tables~\ref{reaction} and \ref{neutron} and
we note that a study of these quantities with two of these structure
models has been made recently \cite{De01} for energies 10 to
300~MeV. The total reaction cross sections for both proton and neutron
scattering from $^{208}$Pb are listed in
Table~\ref{reaction}. Comparing the results of the calculations at
both 40 and 65~MeV with the available proton data indicates a
preference for both SHF, the SKM*, and HO2 models of the
ground state density. However, comparing the model results of the
neutron total reaction cross section with the available evaluated data
at those energies \cite{Ch99} gives a preference for the Skyrme models
only. The predicted proton and neutron total reaction cross sections
at 200~MeV vary sufficiently that their measurements would be
desirable.

The results of our calculations for the total neutron cross sections
are given for the three energies in Table~\ref{neutron} and are
compared with the data of Finlay {\em et al.} \cite{Fi93}. For 40~MeV
the Skyrme models are preferred, although all results overestimate the
measured value. At 65~MeV the Skyrme model results agree well with the
data, the SKM* model result doing best of all, and do better than both
HO predictions. This is not the case at 200~MeV, where all
results predict the measured value reasonably well. While these total
reaction and total cross section results together indicate a
preference for the Skyrme models, we need additional evidence. We
consider then the angular distributions of each scattering.

\begin{figure}
\centering\epsfig{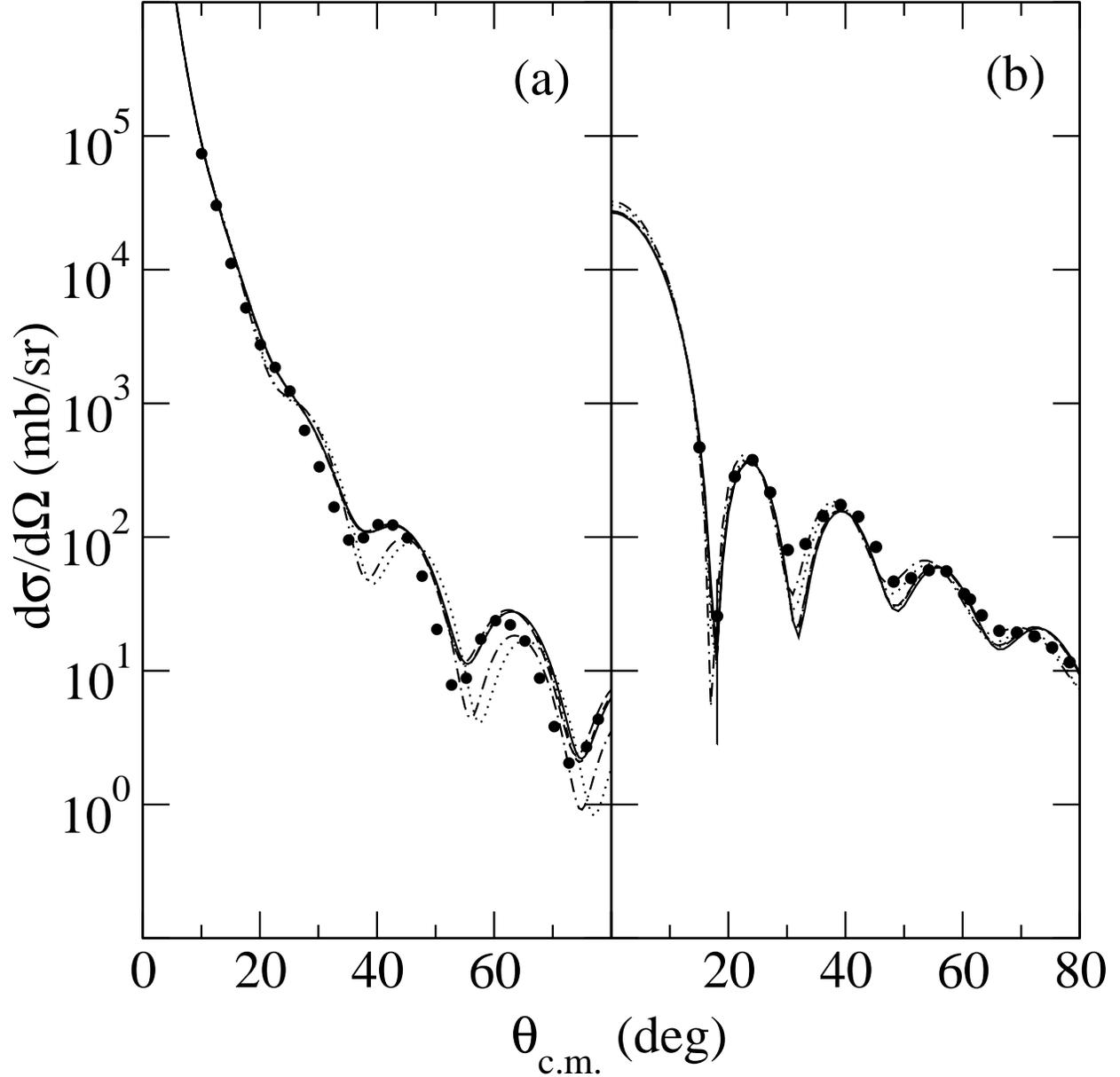}
\caption[]{Differential cross sections for 40~MeV nucleon elastic
scattering from $^{208}$Pb. The proton scattering data of Blumberg
{\em et al.} \cite{Bl66} are compared in (a) to the results from the
SHF1 and SHF2 models (solid and dashed lines respectively), the SKM*
model (double-dot-dashed line), and to the results of the
HO1 and HO2 models (dot-dashed and dotted lines
respectively). The neutron scattering data of de Vito {\em et al.}
\cite{Vi81} are compared to the results of those models as defined in
(a).}
\label{pb_xsec_40}
\end{figure}

\begin{figure}
\centering\epsfig{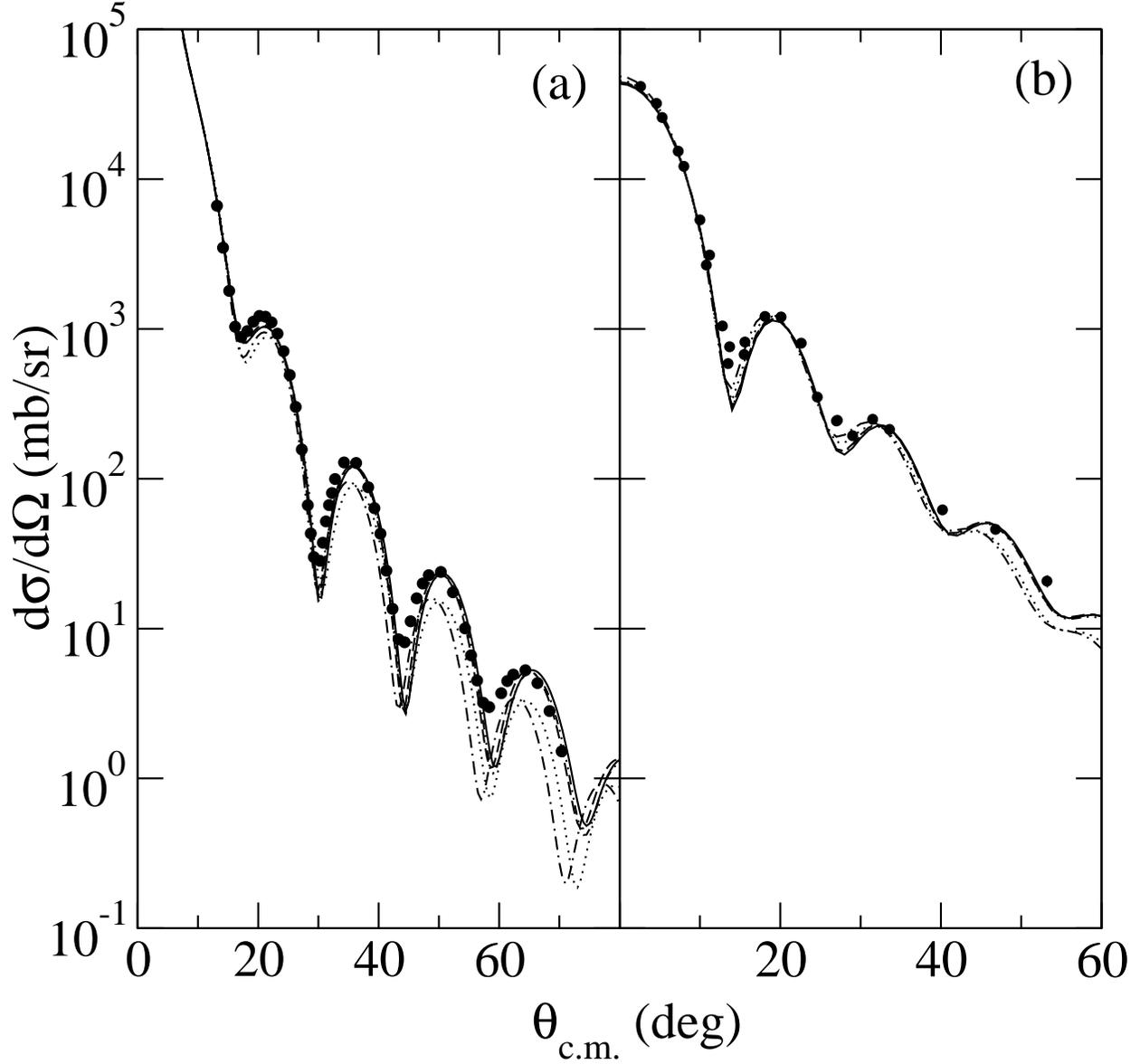}
\caption[]{As for Fig.~\ref{pb_xsec_40}, but for 65~MeV
scattering. The proton scattering data in (a) are those of Sakaguchi
{\em et al.}  \cite{Sa82}, while the neutron scattering data in (b)
are those of Ibaraki {\em et al.} \cite{Ib00}.}
\label{pb_xsec_65}
\end{figure}

\begin{figure}
\centering\epsfig{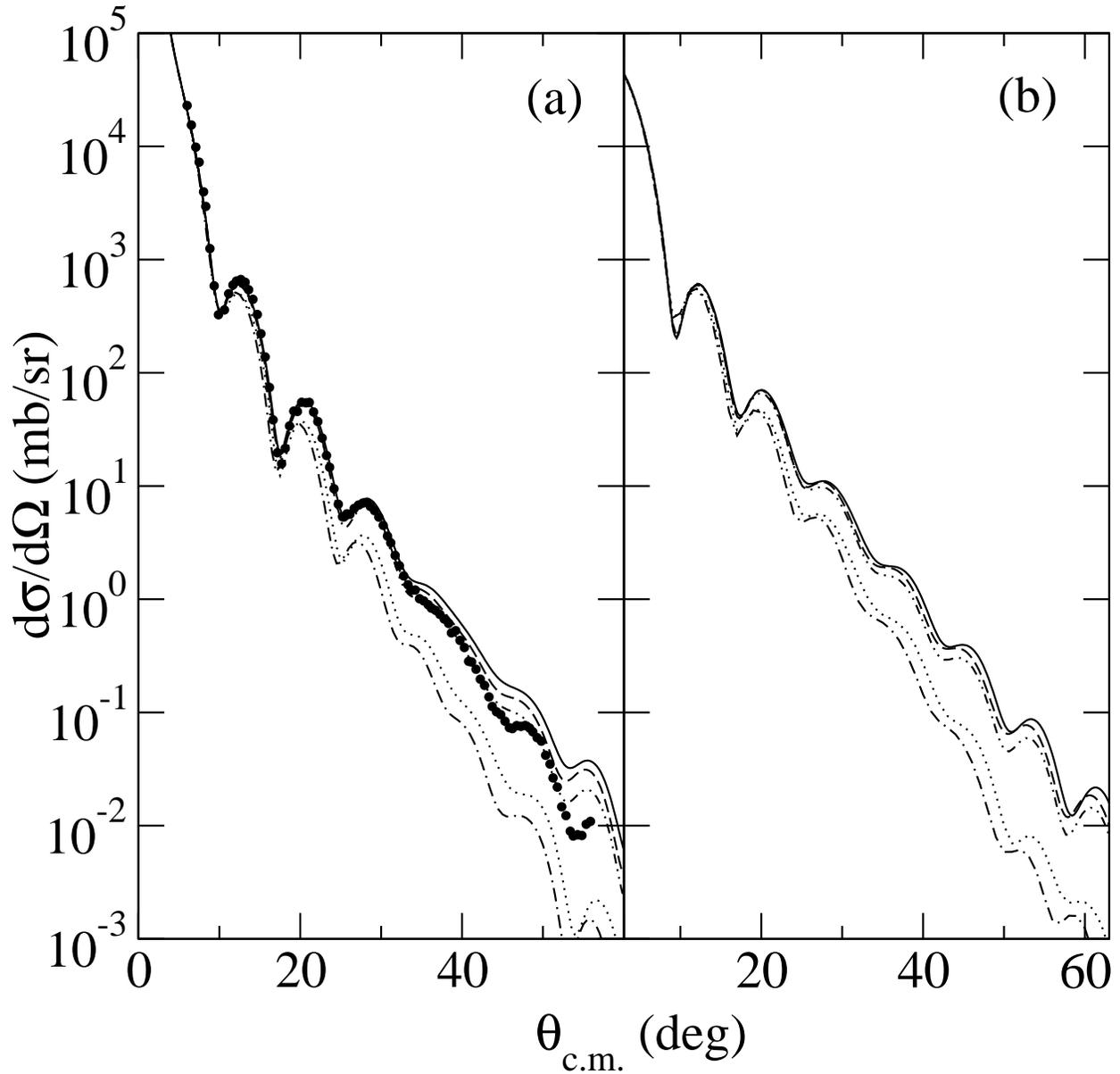}
\caption[]{As for Fig.~\ref{pb_xsec_40} but for 200~MeV scattering. The
proton scattering data are those of Hutcheon {\em et al.} \cite{Hu88}.}
\label{pb_xsec_200}
\end{figure}

The differential cross sections for the scattering of 40, 65 and
200~MeV nucleons from $^{208}$Pb are presented in
Figs.~\ref{pb_xsec_40}, \ref{pb_xsec_65}, and \ref{pb_xsec_200}
respectively.  At 40~MeV, proton elastic scattering is shown in
Fig.~\ref{pb_xsec_40} and evidently there is little if any
differentiation between the SHF1, SHF2, and SKM* model results.  They
both compare well with the data of Blumberg {\em et al.} \cite{Bl66}.
Note, however, that both SHF and the SKM* calculated results agree
much better with the data than do those found using the HO
models of structure.  The quality of reproduction of the data in this
case is in stark contrast to what we found at 40~MeV with $^{40}$Ca.
If that is to remain a problem with specification of the effective
interaction then it seems to be a nucleus-dependent effect.  In the
case of neutron scattering, the results of all model calculations
agree quite well with the data of de Vito {\em et al.}
\cite{Vi81}. The results for $^{208}$Pb are more distinctive than
those for $^{40}$Ca at this and other energies, but so are the
distinctions between the model proton densities for both nuclei.
However, these results indicate that while the nucleon densities in
$^{208}$Pb are better described by the SHF and SKM* models nucleon
scattering at this energy is largely sensitive to the surface
properties only.  Only in the surface are the proton densities still
sufficiently similar for neutron scattering results of all models to
be as alike as they are.

Recall that the integral observables given in Table~\ref{reaction}
make preference to the SHF and SKM* models of the density. Similar
preference is indicated by the differential cross sections for 65~MeV
scattering that are displayed in Fig.~\ref{pb_xsec_65}. Considering
the neutron scattering first, all models give similar results in good
agreement with the data of Ibaraki {\em et al.} \cite{Ib00}, although
above $40^{\circ}$ the SHF and SKM* model results clearly do
better. For proton scattering, the SHF and SKM* models are both in
good agreement with each other and with the data and differ only
slightly from those of the HO models. As for 40~MeV
scattering, the integral observables at 65~MeV concur with these
findings that favor the Skyrme models of the density.

The differential cross sections for 200~MeV nucleon scattering from
$^{208}$Pb are shown in Fig.~\ref{pb_xsec_200}. In this case only
200~MeV proton scattering data exist \cite{Hu88} for comparison with
our predictions.  Nevertheless, that comparison confirms the findings
from analyses of the lower energy data, namely that the SHF and SKM*
models are the better descriptions of the densities of $^{208}$Pb.  In
addition, however, some discrimination between the three Skyrme model
results is evident at this energy. Of the three Skyrme models, the
SKM* model result agrees best of all with the data while the SHF1 does
worst. Given that the rms radii from these two models are very
similar, the marked difference indicates a sensitivity to the
diffuseness in the density. At this energy, unlike those at lower
energies, the neutron scattering results show marked differences
between the SHF and HO calculations.  Again, we expect that this is
due to the scattering of the higher energy probe being more influenced
by the bulk nuclear medium properties of the densities.  Consequently,
a measurement of the 200~MeV neutron elastic scattering, angular
distribution and associated integral observables, is certainly
desirable. 

\section{Conclusions}

The distinctions between the predictions of nucleon scattering from
${}^{208}$Pb at these three energies, and found with the set of five
model structures used, suffice to select that which most likely
prescribes the actual matter densities of the nucleus. A similar
conclusion is reached for $^{40}$Ca for the models considered herein.
Specifically we contend that the use of $g$-folding potential model
calculations can differentiate between different model structures of
the neutron density so as to pin down the neutron rms radius far
better than has been possible in the past. As well the process gives a
good appraisal of the actual density distribution.  For $^{208}$Pb in
particular, our analyses indicate that the SKM* model gives the best
representation of the density.  Together with analyses of the
longitudinal elastic electron scattering form factor it suggests a
neutron skin thickness for $^{208}$Pb of 0.17~fm; a value consistent
with expectations of the SHF1 model nucleus, which is constrained by
the FP neutron equation of state. The only difference between the two
models is that of a larger diffuseness for the SKM* model, accounting
for both the agreement between the results and data for both nucleon
and electron scattering. This would suggest a need to extend the SKX
models to predict a larger diffuseness, for example, by the addition
of a $d$-wave term in the Skyrme Hamiltonian.

One should also note that while a measurement of the skin thickness as
proposed in the experiment for the Jefferson Laboratory~\cite{Je00} is
important, that quantity is a volume property of the nucleon
distributions.  Other information is required to specify a more
complete picture of the neutron density.  From our studies it seems
that simultaneous analyses of angular and integral observables are
relevant.  Given that the HO2, SHF1, and SKM* models of
structure we have used predict essentially the same skin thickness for
$^{208}$Pb but give significantly different predictions when used to
generate $NA$ $g$ folding optical potentials for the nucleon
scattering, as well as electron scattering form factors, analyses of
complementary nucleon and electron scattering data permit one to
discern such finer details of densities.

This work was supported in part by a grant from the Australian
Research Council, U.S. National Science Foundation grants
no. PHY-9605207 and PHY-0070911, and DOE Contract
no. W-7405-ENG-36. Two of us (S.K. and K.A.)  would like to thank the
National Superconduncting Cyclotron Laboratory of Michigan State
University for their kind hospitality during part of this work, while
another one of us (K.A.) would like to thank the Theoretical Division
of the Los Alamos National Laboratory for its similar kind
hospitality.

\bibliography{lead}

%
%	TABLES
%

\begin{table}
\caption[]{Root-mean-square radii (in fm) for protons ($r_p$) and
neutrons ($r_n$) in $^{208}$Pb. The models are as defined in the text.}
\label{radii}
\begin{tabular}{cccc}
Model & $r_p$ & $r_n$ & $r_n - r_p$ \\
\hline
HO1 & 5.45 & 5.83 & 0.38 \\
HO2 & 5.45 & 5.61 & 0.16 \\
SHF1 & 5.45 & 5.61 & 0.16 \\
SHF2 & 5.45 & 5.70 & 0.25 \\
SKM* & 5.45 & 5.62 & 0.17 \\
\end{tabular}
\end{table}

\begin{table}
\caption[]{Total reaction cross sections (in barn) of nucleon
scattering from $^{208}$Pb. The models used are as specified in the
text.}
\label{reaction}
\begin{tabular}{ccccccc}
 & \multicolumn{2}{c}{40 MeV} & \multicolumn{2}{c}{65~MeV} & \multicolumn{2}{c}{200~MeV} \\
Model & proton & neutron & proton & neutron & proton & neutron \\
\hline
HO1 & 2.07 & 2.69 & 2.11 & 2.32 & 1.79 & 1.77 \\
HO2 & 1.95 & 2.62 & 2.00 & 2.27  & 1.71 & 1.73 \\
SHF1 & 1.89 & 2.51 & 1.99 & 2.19 & 1.68 & 1.69 \\
SHF2 & 1.95 & 2.55 & 2.03 & 2.22 & 1.72 & 1.71 \\
SKM* & 1.92 & 2.54 & 2.01 & 2.21 & 1.69 & 1.70 \\
Expt & $2.01 \pm 0.04$ \cite{Ca75} & 2.50 \cite{Ch99} &
$2.02 \pm 0.06$ \cite{In99} & 2.20 \cite{Ch99} & &  \\
\end{tabular}
\end{table}

\begin{table}
\caption[]{Total cross sections (in barn) of neutron scattering from
$^{208}$Pb. The models used are as specified in the text.}
\label{neutron}
\begin{tabular}{cccc}
Model & 40~MeV & 65~MeV & 200~MeV\\
\hline
HO1 & 5.10 & 4.86 & 3.04 \\
HO2 & 4.94 & 4.72 & 2.97 \\
SHF1 & 4.63 & 4.61 & 2.94 \\
SHF2 & 4.71 & 4.67 & 2.94 \\
SKM* & 4.69 & 4.63 & 2.96 \\
Expt \cite{Fi93} & $4.392 \pm 0.001$ & $4.634 \pm 0.001$ & $2.990 \pm 0.003$ \\
\end{tabular}
\end{table}

%
%	FIGURES
%

%
%	Ca figures:
%

%
%	Pb figures:
%

\end{document}